\pdfoutput=1
\documentclass[12pt,a4paper]{article}

\usepackage[utf8]{inputenc}
\usepackage{amsmath}
\usepackage{amsfonts}
\usepackage{color}
\usepackage{graphicx}
\usepackage{natbib}
\usepackage{fullpage}
\usepackage[affil-it]{authblk}

\DeclareMathOperator{\ld}{ld}

\begin{document}
\title{Binary Willshaw learning yields high synaptic capacity for long-term familiarity memory}
\author{Jo\~{a}o Sacramento\thanks{Electronic address: \texttt{joao.sacramento@ist.utl.pt}} }
\author{Andreas Wichert}
\affil{INESC-ID and Instituto Superior T\'{e}cnico, TU Lisbon,\\
       Av. Prof. Dr. An\'{i}bal Cavaco Silva, 2744-016 Porto Salvo, Portugal}
              
\date{March 15, 2012}

\maketitle

\begin{abstract}
In this work we investigate from a computational perspective the efficiency of the Willshaw synaptic update rule in the context of familiarity discrimination, a binary-answer, memory-related task that has been linked through psychophysical experiments with modified neural activity patterns in the prefrontal and perirhinal cortex regions. Our motivation for recovering this well-known learning prescription is two-fold: first, the switch-like nature of the induced synaptic bonds, as there is evidence that biological synaptic transitions might occur in a discrete stepwise fashion. Second, the possibility that in the mammalian brain, unused, silent synapses might be pruned in the long-term. Besides the usual pattern and network capacities, we calculate the synaptic capacity of the model, a recently proposed measure where only the functional subset of synapses is taken into account. We find that in terms of network capacity, Willshaw learning is strongly affected by the pattern coding rates, which have to be kept fixed and very low at any time to achieve a non-zero capacity in the large network limit. The information carried per functional synapse, however, diverges and is comparable to that of the pattern association case, even for more realistic moderately low activity levels that are a function of network size.

\end{abstract}

\section{Introduction}
Observations of psychophysical and neurophysiological order have brought into attention the so-called familiarity discrimination or detection task, where tested subjects need only to recognise once-seen objects without being asked to recollect detailed feature or context descriptions \citep{Xiang1998,Xiang2004,Yakovlev2008}. From the computational perspective, the essential aim is to devise a neural network model that is biologically plausible up to a certain degree of realism and that is able to explain in part the seemingly limitless memorising ability of the brain to solve this task \citep{Standing1973}.

As in previous familiarity memory neural network modelling efforts \citep{Bogacz2001,Greve2009,Cortes2010}, the formulation of the task that we consider involves a set of $M$ patterns
\begin{equation}
\mathcal{S} = \{\mathbf{x}^1, \ldots, \mathbf{x}^\mu, \ldots \mathbf{x}^M\},
\end{equation}
that have been presented to the network for learning and that ought to be recognised as familiar in future presentations, while any other pattern not belonging to $\mathcal{S}$ should be classified as novel. Each of the patterns is a binary vector $\mathbf{x}^\mu \in \{0, 1\}^m$, $x_i^\mu$ representing the (silent-firing) activity of the $i$-th neuron at a given time frame $\mu$; the task itself is as well binary, in the sense that we seek to decide if a certain presented pattern $\tilde{\mathbf{x}}$ is either familiar or novel. The structure of the network is given at any time by the $m \times m$ connectivity matrix $\mathbf{W}$, where the entry $w_{ij}$ denotes the strength of the bond from presynaptic neuron $i$ to postsynaptic neuron $j$.

To learn the desired mapping, each neuron should be able to determine at the synapse level (`locally') the network connectivity structure so that in subsequent pattern presentations one can extract from the collective activity of the $m$ neurons the desired novel-familiar response. The model is then characterised by a local synaptic learning rule and by a discrimination function. On the one hand, given a pattern $\mathbf{x}^\mu$ that should be memorised, the former determines each synaptic weight solely by inspection of the variables $w_{ij}$, $x_i$ and $x_j$; the latter, given a query pattern $\tilde{\mathbf{x}}$ and the structure of the network $\mathbf{W}$, elicits the binary familiarity response.

We focus on modelling long-term memory, in opposition to palimpsestic working memory \citep{Parisi1986,Amit1994,Leibold2008,Barrett2008,Yakovlev2008}, where `overwriting' takes place and the familiarity signal of past memories decays over time. For long-term familiarity detection, a model that is capable of storing an extensive number of patterns per synapse has been proposed \citep{Bogacz2001} and recently shown to correspond to the optimal linear, local familiarity learning prescription \citep{Greve2009}. However, the network is only capable of storing a rather small amount of information per synapse, and the proposed synaptic update scheme requires maintenance of real-valued synapses over a long period of time.

In our work, we consider as an alternative the binary non-linear Willshaw (or Steinbuch) prescription \citep{Steinbuch1961a,Willshaw1969} in the context of familiarity discrimination. This learning rule has certain properties that have made it desirable when applied to the associative memory problem, where it has been extensively analysed \citep[see, e.g.,][]{Willshaw1969, Palm1980, Golomb1990, Nadal1990, Palm1992, Buckingham1992, Brunel1994, Graham1995, Sommer1999, Knoblauch2010}; namely, the high storage capacity attained when the model is correctly parametrised, its simplicity, and the fact that the generated synaptic matrix $\mathbf{W}$ is binary. This last feature is particularly interesting since in cortical regions supporting memory-related tasks the synaptic transitions may operate in a discrete (few steps) or even in a binary switch-like fashion. There is accumulating experimental evidence supporting discrete transitions at least in the initial phase of long-term potentiation, although it remains unclear whether or not long-term synaptic efficacies may still have a gradual distribution \citep{Petersen1998,Montgomery2004,OConnor2005}.

Furthermore, an inhibitory variant of the Willshaw rule has just been proposed by \citet{Knoblauch2010}, motivated by the possibility of structural plasticity by synaptic pruning and growth as a support for long-term memory encoding in the adult mammalian brain \citep{Chklovskii2004}, alongside well-established synaptic weight change mechanisms such as long-term potentiation and depression. In the associative case, the inhibitory Willshaw rule has led to the discovery of new efficient working regimes where few active synapses can carry a high Shannon information content.

In this article we show in a first step that for medium-sized networks the classical pattern and Shannon capacities of the Willshaw model are comparable to those of the real-valued network of \citet{Bogacz2001}, provided that the patterns exhibit low activity levels at any time (the so-called sparse coding regime), a fact that has already been pointed out in the dynamical synapse analysis of \citet{Barrett2008}. We also show that in the limit of large networks $m\to \infty$, the network capacity vanishes unless the coding rates are extremely low.

In line with the recent observations of \citet{Knoblauch2010}, we then investigate alternative parametrisations of the Willshaw model. We find that the high pattern loadings associated with the familiarity discrimination task lead to dense potentiation of the memory matrix, a regime where the inhibitory interpretation of the original Willshaw model is especially efficient. It is shown that if the low cost of silent synapses (which might even be pruned in the long-term) is neglected, the inhibitory network is capable of achieving large synaptic capacities that increase with the number of neurons, under realistic moderately low coding rates. Finally, we take into consideration the effects of varying the coding level per pattern; at least when the level follows a binomial distribution, introducing a feedforward inhibitory correction in the discriminator compensates for the additional signal variability and the system remains qualitatively intact, albeit operating with lower overall efficiency in the finite-size case.

\section{Results}
The simplest possible local, non-linear, binary synaptic rule is the well-known Willshaw prescription \citep{Steinbuch1961a, Willshaw1969, Palm1980}. Here, the weight update equation is an extreme case of Hebbian learning, where a single coincidental firing activity at any given time $\mu$ (i.e., $x^\mu_i=1$ and $x^\mu_j=1$) is sufficient to arise long-term potentiation at the synaptic contact $i \to j$. As there is just one potentiation level, each synapse $w_{ij}$ is a binary variable, either at the 0-state (silent synapse) or at the 1-state (present synapse). After $M$ pattern presentations, $w_{ij}$ is given by
\begin{equation}
\label{eq:Willshaw-learn}
w_{ij} = \min \left(1, \sum_{\mu=1}^M x_i^\mu x_j^\mu \right) \in \{0, 1\}.
\end{equation}

Originally proposed in the context of an associative network with one-step (non-iterative) synchronous retrieval, the 0-1 Hebb rule \eqref{eq:Willshaw-learn} has been employed as well to embed patterns in attractor networks with symmetric couplings $w_{ij}=w_{ji}$. In this case, if an appropriate retrieval strategy is used so as to form large basins of attraction surrounding the desired fixed points, iteration generally leads to a more robust recall process, in terms of allowed cue distortion (given by a metric such as the Hamming distance $d_H(\tilde{\mathbf{x}}, \mathbf{x}^\mu) \equiv \sum_i |x^\mu_i - \tilde{x}_i|$) as well as in terms of resistance to stochastic synaptic failure, where the $w_{ij}$ may randomly switch states with a certain probability \citep{Golomb1990,Schwenker1996a,Sommer1998a}.

For familiarity discrimination, there is no need per se to extract the whole pattern $\mathbf{x}^\mu$ from the network; rather, what one seeks is a prescription to determine a binary (novel-familiar) answer starting from a cue $\tilde{\mathbf{x}}$, given the information stored in the synaptic connectivity matrix $\mathbf{W}$.

The discriminator proposed by \citet{Bogacz2001} and studied in formal memory models of familiarity \citep{Bogacz2003a,Greve2010}, is based on the quadratic form
\begin{equation}
\label{eq:energy}
H(\mathbf{x})=-\alpha \mathop{\sum \sum}_{i \neq j} w_{ij} (x_i-f) (x_j-f) \in \mathbb{R},
\end{equation}
usually referred to as the energy function\footnote{As for bipolar patterns and symmetrical networks ($w_{ij}=w_{ji}$) with no self-couplings ($w_{ii}=0$) there is a strong analogy with the Hamiltonian of the zero-temperature Ising model \citep{Hopfield1982,Amit1985a}.} of the network at a given state $\mathbf{x}$, presented in its mean corrected form \citep{Amit1987,Bogacz2002,Greve2009}, where $f \equiv m^{-1} \mathrm{E}(\sum_i x_i)$ is the coding rate, i.e., the expected fraction of firing units per pattern. As it has already been pointed out in the previous works, equation \ref{eq:energy} has a network implementation and it is closely related to other measures of familiarity \citep[see, e.g., the appendix of][]{Greve2010}.

In the proposed discrimination scheme, the desired binary decision is computed by `clamping' into the network state a certain input pattern $\tilde{\mathbf{x}}$ and then, without (or before) the retrieval dynamics takes place, by thresholding the resulting energy, i.e.
\begin{equation}
\label{eq:discriminator}
D(\tilde{\mathbf{x}})=\mathbf{1}_{[H(\tilde{\mathbf{x}}) \le \Theta]} \in \{0, 1\},
\end{equation}
where $\mathbf{1}_{[\cdot]}$ is the binary random variable which is 1 if the argument holds and 0 otherwise. An appropriate choice of $\alpha$ and $\Theta$ should ensure that, given a weight matrix $\mathbf{W}$ encoded according to a certain synaptic learning rule, as many as possible patterns belonging to $\mathcal{S}$ are assigned one of the two decision outcomes (say, one), and all the others to the opposite class (say, zero).

It has been recently shown by \citet{Greve2009} that for such discriminator, the asymptotically optimal ($m \to \infty$ and a size-dependent load $M$) local linear synaptic weight setting when we allow the $w_{ij}$ to assume real values is given by the covariance learning rule \citep{Amit1987,Tsodyks1988,Dayan1991,Palm1996}:
\begin{equation}
\label{eq:linear-hebb}
w_{ij} \propto \sum_{\mu=1}^M (x_i^\mu-f) (x_j^\mu-f) \in \mathbb{R}.
\end{equation}

In this article we address the question of how well does the clipped Hebbian rule \eqref{eq:Willshaw-learn} fare with a discriminator of the form \eqref{eq:discriminator}. Specifically, for simplicity we redefine $H$ letting $\alpha=1$, performing the double summation over all $i,j$, and dropping the mean correction,
\begin{equation}
H(\mathbf{x})=-\sum_{i=1}^m \sum_{j=1}^m w_{ij} x_i x_j \in \mathbb{Z},
\end{equation}
recalling that each weight $w_{ij}$ is now a 0-1 binary variable.

Following the analysis of the associative Willshaw network carried out by \citet{Knoblauch2010}, we proceed by calculating three essential quantities: the maximal number of patterns $M_{\epsilon}$ that the system can discriminate allowing a certain (known) error level, the network capacity $C$ (in bits per synaptic contact), and the synaptic capacity $C^S$ (in bits per active synapse). We will then see that the Willshaw model becomes especially interesting regarding the latter quantity, as a modification to the clipped rule leads to the activation of a subset of few synapses within the full contact space of order $m^2$.

\subsection{Maximal pattern load calculation for low activity levels}
\label{sec:pattern-capacity}
The calculation of the maximal pattern load $M_{\epsilon}$ when the average activity is low ($f \ll 1$) can be performed analytically using a series of approximations which have been shown to be near-exact even for finite networks where $m$ is not large \citep{Palm1980,Knoblauch2008,Knoblauch2010}.

We consider the two usual simplified binary pattern generation scenarios: first, we deal with the case where every pattern $\mathbf{x}^\mu$ presented to the network for learning has a fixed, known a priori activity level $|\mathbf{x}^\mu| \equiv \sum_{i=1}^m x_i = k$ as in the analysis of \citet{Palm1980}; later (in section \ref{sec:binomial}), we consider patterns where $|\mathbf{x}^\mu|$ is a binomially-distributed random variable with characteristic probability equal to the coding rate $f=k/m$, $k$ being again a fixed known a priori parameter. In this case, although the activity of each pattern is allowed to vary, by construction the average level is $mf=k$ and all neurons are activated equally and independently \citep{Buckingham1992}.

With these statistics at hand we can determine the average weight matrix load,
\begin{align}
p_1 &\equiv \mathrm{E}(w_{ij}) = \mathrm{P}(w_{ij}=1)=1-\mathrm{P}(w_{ij}=0)\\
&=1-(1-f^2)^M = 1-\exp(M \ln (1-f^2))\label{eq:p1}\\
&\approx 1-\exp(-f^2M).\label{eq:p1approx}
\end{align}
The approximation assumes that the coding rates are low, i.e., $f^2 \ll 1$.

Clearly, as observed when employing the Willshaw rule to solve the associative task, $p_1$ is a critical quantity: to recover information about the patterns in $\mathcal{S}$ one must control both the cardinality $M$ and the sparseness parameter $f$ so as to avoid $p_1=1$. It is useful to calculate $M$ given $p_1$,
\begin{equation}
\label{eq:M-by-p1}
\ln(1-p_1) \approx -Mf^2 \Leftrightarrow M \approx -f^{-2} \ln(1-p_1).
\end{equation}

Regarding familiarity detection in general, two types of error may occur: omission errors (denoted as `10' errors) whenever $\tilde{\mathbf{x}} \in \mathcal{S}$ but the system fails to classify the pattern as familiar; conversely, commission errors (denoted as `01' errors) when $\tilde{\mathbf{x}} \notin \mathcal{S}$ but the discriminator indicates familiarity. For patterns with fixed (for all $\mu$) activity $k$ and $\mathbf{W}$ set according to the Willshaw rule \eqref{eq:Willshaw-learn}, there is a simple threshold setting which avoids omission errors at all, i.e., a $\Theta$ such that for all $\mu$ we have with probability one $D(\mathbf{x}^\mu)=1$. For a familiar cue $\tilde{\mathbf{x}} \in \mathcal{S}$ corresponding to a certain learned $\mathbf{x}^\mu$ we have
\begin{align}
\label{eq:ThetaW-familiar-energy}
H(\tilde{\mathbf{x}}) &= - \sum_{i=1}^m \sum_{j=1}^m w_{ij} \tilde{x}_i \tilde{x}_j\\
&= - \sum_{i=1}^m \sum_{j=1}^m x^\mu_i x^\mu_j = - k^2 \equiv \Theta_W,
\label{eq:ThetaW}
\end{align}
where the equality from \eqref{eq:ThetaW-familiar-energy} to \eqref{eq:ThetaW} is valid since $w_{ij}=1 \Leftrightarrow \exists \mu,\; x_i^\mu=1 \wedge x_j^\mu=1$. In a sense, $\Theta_W$ is the familiarity discrimination threshold which corresponds to the classical Willshaw threshold $|\tilde{\mathbf{x}}|=k$ for the noise-free associative task \citep{Willshaw1969,Palm1980}.

When $\Theta_W$ is the discrimination threshold and $\tilde{\mathbf{x}}$ is a novel pattern, generated according to the same statistics as the $\mathbf{x}^\mu$ but not presented for learning, if the non-zero $w_{ij}$ coincide with active $i,j$ units enough such that $H(\tilde{\mathbf{x}})$ reaches $-k^2$, a commission error will occur. We can calculate this error probability resorting to $p_1$; assuming that the `ones' in $\mathbf{W}$ were randomly and independently set\footnote{A well-known approximation employed e.g. in the analyses of \citet{Willshaw1969,Palm1980,Knoblauch2010}, which is valid for sparse patterns with activity levels that are sublinear in $m$ \citep{Knoblauch2008}.},
\begin{align}
p_{01} &\equiv \mathrm{P}(D(\tilde{\mathbf{x}})=1 \mid \tilde{\mathbf{x}} \notin \mathcal{S}) \approx \mathrm{P}(D(\tilde{\mathbf{x}})=1)\\
&\approx {p_1}^{(-\Theta_W-k)/2}\label{eq:p01-k}\\
&\approx {p_1}^{k^2/2},
\label{eq:p01}
\end{align}
where the $1/2$ correction comes from the symmetry in $\mathbf{W}$. To reach our final expression \eqref{eq:p01}, we approximate $(k^2-k)/2$ by the leading term $k^2/2$, although equation \ref{eq:p01-k} would yield a better approximation to the true value of $p_{01}$ as the learning rule \eqref{eq:Willshaw-learn} sets the diagonal entries of $\mathbf{W}$ to one with high probability.

While parametrising a memory device, to ensure the system performs the desired task correctly it is common to require that the probability of error remains below a certain bound. In the associative memory literature there are many criteria to enforce a quality level in the process; usually, the task parameters are found so that the error probability grows according to some controlled function of network size and the expected pattern activity level \citep{Palm1980,Knoblauch2010}. In the familiarity detection task, however, as there is no obvious reason to couple the probabilities to the parameters $k$ and $m$, it seems reasonable to maintain $p_{01}$ and $p_{10}$ below a fixed level \citep{Bogacz2002}.

To keep the error probability $p_{01}$ lower than a desired level $p_{01\epsilon}$, we establish the `breakdown' value $M_\epsilon$ for the pattern load, as a function of the coding rate $f$. Using the binomial approximation given by equation \ref{eq:p01}, we have
\begin{equation}
\label{eq:p01eps}
p_{01} \approx p_{01\epsilon} \Leftrightarrow \left(1-\exp\left(-f^2M\right)\right)^{k^2/2} \approx p_{01\epsilon},
\end{equation}
yielding, with respect to $M$,
\begin{align}
M &\approx -\frac{m^2}{k^2} \ln\left(1-{p_{01\epsilon}}^{2/k^2}\right) \equiv M_\epsilon,
\label{eq:Meps}
\end{align}
which is the pattern capacity we sought. Note that in the large network limit $m\to\infty$, for any coding rate such that $k \to \infty$, $M_\epsilon$ is independent of the fixed error bound $p_{01\epsilon}$, as we have
\begin{equation}
\label{eq:Meps-limit}
M_\epsilon \approx \frac{2 m^2 \ln k}{k^2}.
\end{equation}

Notice how the maximal pattern load is a function of $k$ and $m$. This result is in contrast with the real-valued network employing the covariance rule, where the familiarity discrimination capacity is essentially independent of the pattern activity level \citep{Bogacz2002}. Just as in the analyses of the Willshaw rule for the associative case \citep{Willshaw1969,Palm1980,Nadal1990,Knoblauch2010}, however, we find a dependence of $M_\epsilon$ on $k$. With the binary synapses induced by Willshaw learning, it is clear that $M_\epsilon$ is maximised in the sparse coding regime $f \ll 1$; the actual optimal activity level parameter $k^\text{opt}$ is just a function of $p_{01\epsilon}$ and can easily be found numerically. To gain additional insight on the typical size of $k^\text{opt}$, let us obtain an approximation for the pattern capacity,
\begin{equation}
\label{eq:Meps-expanded}
M_\epsilon \approx \frac{m^2}{k^2} \left(2\ln k - \ln\left(-2 \ln p_{01\epsilon}\right)\right),
\end{equation}
which is maximal when
\begin{equation}
\label{eq:kopt}
k = \exp\left(\frac{1}{2}\left(1+\ln\left(-2 \ln p_{01\epsilon}\right)\right)\right) \approx k^\text{opt}.
\end{equation}
Recalculating $M_\epsilon$ with $k=k^\text{opt}$, we find that
\begin{align}
\label{eq:max-Meps}
\max_{k} M_\epsilon &\approx -\frac{1}{2 e \ln p_{01\epsilon}} m^2 \\
&\approx 0.18 (- \ln p_{01\epsilon})^{-1} m^2.
\end{align}
Just to illustrate the result above, if one sets the desired error rate at $p_{01\epsilon}=0.01$, the obtained breakdown quantity of patterns per synapse becomes about $M_\epsilon/m^2 \approx 0.04$.

Although `greedily' maximising $M_\epsilon$ leads to an extensive quantity of patterns per synapse, this approach also imposes a heavy coding restriction in the form of quite small values for $k$ and an optimising expression that does not vary with $m$, a parametrisation that is referred to by \citet{Knoblauch2010} as the ultra-sparse coding regime. In the next sections we proceed to richer performance measures where the required underlying resources and the Shannon information of the task are also taken into account.

\subsection{Classical network capacity}
The commission error probability $p_{01}$ can as well be used to calculate the traditional network capacity measure $C$ in bits per synaptic contact. Here there is a fundamental difference between the associative and familiarity tasks, as observed by \citet{Barrett2008,Greve2009}: a familiarity discrimination network can only `transmit' at most one bit per learned pattern (the perfect output of $D(\tilde{\mathbf{x}})$), instead of order $k$ bits per pattern as in the associative case \citep{Palm1980,Knoblauch2010}. The optimal local, linear, additive covariance rule (that induces real-valued synaptic weights) can then only obtain $0.057$ bits per synapse in the $M\to\infty$ errorful regime \citep{Greve2009}, which is rather low when compared to the $0.72$ bits per synapse that the same rule can achieve in the high fidelity pattern association task \citep{Palm1996}.

The analogy at hand is to interpret the familiarity network as a discrete binary channel which transmits novel and familiar patterns with a certain error probability, and then calculate the information-theoretic channel capacity, which is the maximal mutual information \citep{Shannon1948,Cover2006} normalised by the number of required synaptic contacts,
\begin{align}
\label{eq:network-capacity}
C = \frac{I(X^1, \ldots, X^\omega, \ldots, X^\Omega; Y^1, \ldots, Y^\omega, \ldots, Y^\Omega)}{m^2}.
\end{align}
Here $X^\omega \in \{0, 1\}$ is a binary random variable indicating whether the $\omega$-th presented pattern is familiar ($X^\omega=1$) or novel ($X^\omega=0$), and $Y^\omega \equiv D(\mathbf{x}^\omega) \in \{0, 1\}$ is the network output for the $\omega$-th pattern. As in previous work \citep{Barrett2008,Greve2009}, we assume that $\Omega=2M$ patterns are presented and an equal prior probability of a pattern being familiar or novel $P(X^\omega=0)=P(X^\omega=1)=1/2$. Besides allowing for a direct fair comparison with the previously obtained results, a prior model with equiprobable pattern classes maximises the channel capacity when the conditional error probabilities are equal $p_{10}=p_{01}$. In our case, assuming the network is parametrised for high fidelity, this choice is approximately optimal, as we have $p_{10}=0$ and $p_{01} \approx 0$.

Since we are `transmitting' $M$ learned and $M$ novel patterns independently generated according to the statistics of section \ref{sec:pattern-capacity}, the process can be decomposed into $2M$ transmissions of a single (say, the $\omega$-th) pattern,
\begin{align}
C &= \frac{2M}{m^2} I(X^\omega; Y^\omega)\\
&= \frac{2M}{m^2}\Bigg[1-\frac{1}{2}\Big((1+p_{01}) \ld(1+p_{01})-p_{01} \ld p_{01}\Big)\Bigg],
\label{eq:network-capacity-2}
\end{align}
where $p_{01}$ is the commission error probability, defined in \eqref{eq:p01} as a function of the task parameters $m$, $k$, $M$. The derivation of the single-pattern mutual information is given in appendix \ref{apx:mutual-info}; a similar calculation has been carried out in the single-neuron information maximisation framework of \citet{Barrett2008}, in a comparison of the Willshaw rule with more elaborate stochastic synaptic learning.

\begin{figure}[ht]
\begin{center}
\includegraphics[scale=1.2]{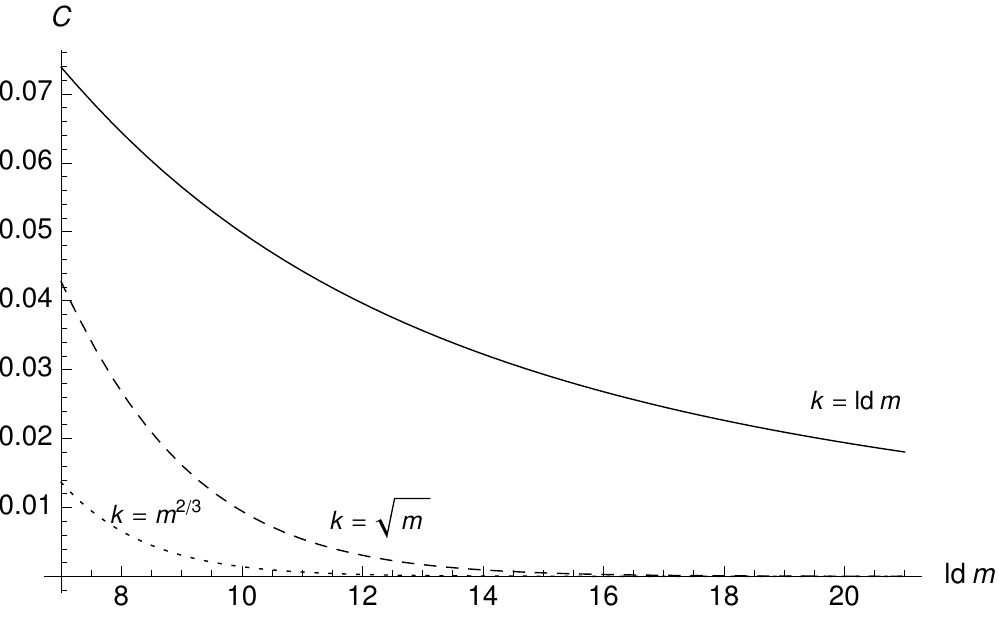}
\caption{Network capacity $C$ in bits per synaptic contact vs. network size $m$ (in logarithmic scale) for a variety of activity level orders, with pattern load $M_{\epsilon}$ given by \eqref{eq:Meps} at conditional error rate $p_{01\epsilon}=0.01$. For $k$ of order $\log m$ the capacity is stable, yet slowly decreasing towards zero as predicted by the asymptotic analysis. Less sparse patterns (e.g., when $k=\sqrt{m}$) lead to low capacity even for small $m$. When the activity level increases to $k=m^{2/3}$ the network capacity becomes near-zero for any network size. For $p_{01\epsilon}=0.01$, integer-constrained numerical optimisation of $C$ with respect to $k$ while $M$ is accordingly set at $M_\epsilon$ reveals that the maximum $C\approx0.11$ is achieved when $k=4$, a result which is in agreement with the previous findings of \citet{Barrett2008}.\label{fig:C-k-m-large}}
\end{center}
\end{figure}

Unfortunately, unlike the network capacity achieved in the associative case, in our task $C$ is largest for finite small $m$ (see figure \ref{fig:C-k-m-large}), but vanishes when $m\to\infty$, for any activity level function $k$ that increases with $m$.

To show this, let us take an arbitrary, finite probability $p_{01\epsilon}$ close to zero, to keep the discrimination error from growing large; in this case, the bracketed quantity in \eqref{eq:network-capacity-2} becomes approximately one. Then, the capacity becomes
\begin{equation}
\label{eq:C}
C \approx  - 2 k^{-2} \ln\left(1-{p_{01\epsilon}}^{2/k^2}\right).
\end{equation}
In the limit $k,m \to \infty$, we can take $M_\epsilon$ from equation \ref{eq:Meps-limit}; the capacity $C$ no longer depends on the error bound $p_{01\epsilon}$ and is given by
\begin{equation}
C \approx \frac{4 \ln k}{k^2}.
\end{equation}

We have reached a result which describes a qualitative behaviour that is rather different from the one found in the typical long-term associative memory task, where capacity is clearly a function of network size, and an increasing one when the activity level $k$ is of correct order \citep{Willshaw1969,Palm1980,Dayan1991}. For a given fixed probability error $p_{01\epsilon}$, the capacity $C$ of the Willshaw network for discrimination is not directly a function of network size $m$. In our case, for any order of $k$ as an increasing function of $m$, in the limit of $m\to\infty$, the capacity of the system collapses, even if the limit is reached slowly. One can avoid near-zero capacity for large networks only in the ultra-sparse regime, where $k$ is kept small and constant (e.g., $k=4$) and the capacity remains non-zero (and independent of $m$).

\subsection{Synaptic capacity}
Let us consider now the synaptic capacity measure $C^S$ (in bits per active synapse) recently suggested by \citet{Knoblauch2010}. Here, only functional synapses (i.e., non-zero synaptic connections $w_{ij}$ which play a role in the network task) are considered to count; silent synapses are either assumed to be wired but metabolically cheap to maintain or even that the network is endowed with structural plasticity and is able to prune irrelevant synapses and rewire new connections as needed \citep[e.g.,][]{Poirazi2001,Chklovskii2004,Holtmaat2009}. In the simple pattern statistics we consider, we obtain $C^S$ by renormalising the network capacity $C$ (as given by equation \ref{eq:network-capacity-2}) by a factor $F$ denoting the fraction of functional synapses:
\begin{equation}
\label{eq:CS}
C^S = \frac{C}{F} = \frac{2M}{F m^2} I(X^\omega; Y^\omega).
\end{equation}

\begin{figure}[ht]
\begin{center}
\includegraphics[scale=1.2]{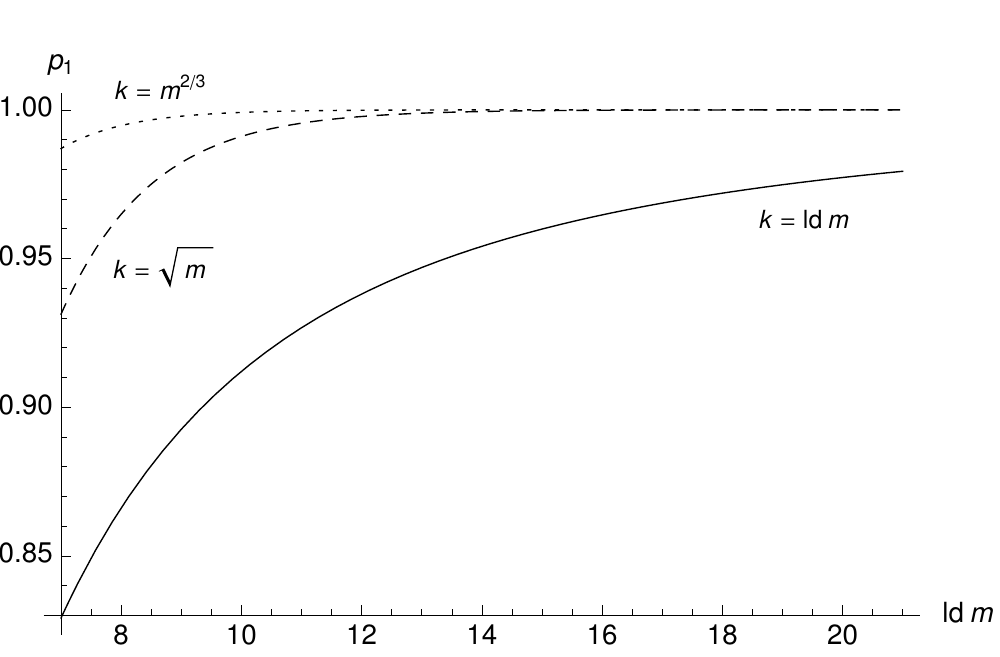}
\caption{The ratio $F\equiv C/C^S=p_1$ between network and synaptic capacities for the Willshaw model, when the error probability bound is $p_{01\epsilon}=0.01$, shown for different activity functions $k(m)$. Since the maximal network capacity for each pair $(m, k(m))$ is achieved at a higher connectivity level $p_1$ as the coding rate increases, the relative advantage of considering only functional synapses becomes negligible.\label{fig:p1}}
\end{center}
\end{figure}

In the classical Willshaw model, the functional elements correspond to the 1-synapses, the expected fraction of which is $p_1$ (our $F$, then) as defined in equation \ref{eq:p1}. However, at the maximal pattern load $M_\epsilon$, even when the discrimination error bound $p_{01\epsilon}$ is kept low, most synapses are in the potentiated state. We can see this by rewriting $p_1$ as a function of $p_{01\epsilon}$; when $M$ is given by $M_\epsilon$, combining equations \eqref{eq:p01} and \eqref{eq:p01eps}, we obtain
\begin{align}
\label{eq:p1-from-p01eps}
p_1 \approx {p_{01\epsilon}}^{2/k^2} \gg 1/2,
\end{align}
which approaches unity as we let $k\to \infty$ and is already larger than $1/2$, even for small $p_{01\epsilon}$ close to zero and low activity $k$. Once again, in the limit $m\to\infty$, when $k$ is allowed to vary as a function of $m$, we have $Fm^2 \to m^2$, which implies a capacity collapse $C^S \to C \to 0$. The differences between $C^S$ and $C$ for finite $m$ are also rather small, as illustrated by figure \ref{fig:p1}.

However, parametrisations leading to the so-called dense potentiation regime $p_1 \to 1$ (as $m \to \infty$) can be quite advantageous in terms of synaptic capacity when the connectivity matrix $\mathbf{W}$ is set according to the inhibitory Willshaw learning rule. In the associative task, this rule is able to achieve a synaptic capacity already an order of magnitude larger than that of the original excitatory model for reasonable pattern activity $k$ and plausible network size, and arbitrarily higher values in large networks with appropriate activity levels \citep{Knoblauch2010}. Furthermore, it is one of the limit cases of the optimal non-linear Bayesian local synaptic update \citep{Knoblauch2011}.

The inhibitory rule is a subtle variation of equation \eqref{eq:Willshaw-learn}, as the synaptic states set by the original rule are simply switched: each 0-synapse (encoding non-coincidental activity) becomes functional as an inhibitory synapse $w_{ij}=-1$; conversely, each $1$-synapse becomes silent $w_{ij}=0$. We denote the synaptic connectivity matrix of the inhibitory variant by $\tilde{\mathbf{W}}$; after $M$ pattern presentations, the state of synapse $i\to j$ is
\begin{equation}
\label{eq:inhibitory-rule}
\tilde{w}_{ij}=w_{ij}-1=\max\left(-1, \sum_{\mu=1}^M x_i^\mu x_j^\mu -1 \right),
\end{equation}
where $w_{ij}$ is the 0-1 weight that would be induced by the excitatory rule.

The energy for a familiar cue $\tilde{\mathbf{x}} \in \mathcal{S}$ is now $\Theta_I \equiv H(\tilde{\mathbf{x}})=0$, following the reasoning which led to the derivation of $\Theta_W$. Novel patterns should activate the inhibitory synapses so that for a given $\tilde{\mathbf{x}} \notin \mathcal{S}$, $H(\tilde{\mathbf{x}}) > 0 = \Theta_I$; thus, the discrimination function \eqref{eq:discriminator} remains unchanged. The (classical) network capacity of the inhibitory network is

Notice that the excitatory and inhibitory networks are functionally equivalent and that the (classical) network capacities of both implementations are equal, i.e., $\tilde{C}=C$. It is the synaptic capacity $\tilde{C}^S$ of the inhibitory network the fundamental quantity to observe, as it is inversely proportional to the fraction $\tilde{F}$ of inhibitory synapses
\begin{align}
\label{eq:F_tilde}
C/\tilde{C}^S&=\mathrm{P}(\tilde{w}_{ij}=-1)=1-p_1=(1-f^2)^M\\
&\approx \exp(-f^2 M) \equiv \tilde{F},
\end{align}
where we have used approximation \eqref{eq:p1approx} for $p_1$.

Alternatively, $\tilde{F}$ can be obtained as a function of the error probability bound $p_{01\epsilon}$ from \eqref{eq:p1-from-p01eps},
\begin{equation}
\label{eq:F_tilde_eps}
\tilde{F} = 1-p_1 \approx 1-{p_{01\epsilon}}^{2/k^2}.
\end{equation}
Expanding the network capacity $C$ as in \eqref{eq:C} and inserting in \eqref{eq:CS} the factor $\tilde{F}$ we have just derived, we arrive at the synaptic capacity of the inhibitory network as a function of $k$ and $p_{01\epsilon}$:
\begin{equation}
\label{eq:CS_p01eps}
\tilde{C}^S \approx - \frac{2 \ln\left(1-{p_{01\epsilon}}^{2/k^2}\right)}{k^2 \left(1-{p_{01\epsilon}}^{2/k^2}\right)},
\end{equation}
which is approximately
\begin{equation}
\label{eq:CS_approx1}
\tilde{C}^S \approx - \frac{2 \ln k - \ln \left(-2 \ln p_{01\epsilon}\right)}{\ln p_{01\epsilon}},
\end{equation}
the approximation improving as $k$ increases.

Asymptotically, letting $k \to \infty$, the capacity further simplifies to
\begin{equation}
\label{eq:CS_approx2}
\tilde{C}^S \approx - \frac{2 \ln k}{\ln p_{01\epsilon}}.
\end{equation}
Notice that for large $k$, the $k^{-2}$ factor that was hampering the capacity in the excitatory model has disappeared, both in the finite case \eqref{eq:CS_approx1} and in the large network limit \eqref{eq:CS_approx2}.

\begin{figure}[ht]
\begin{center}
\includegraphics[scale=1.2]{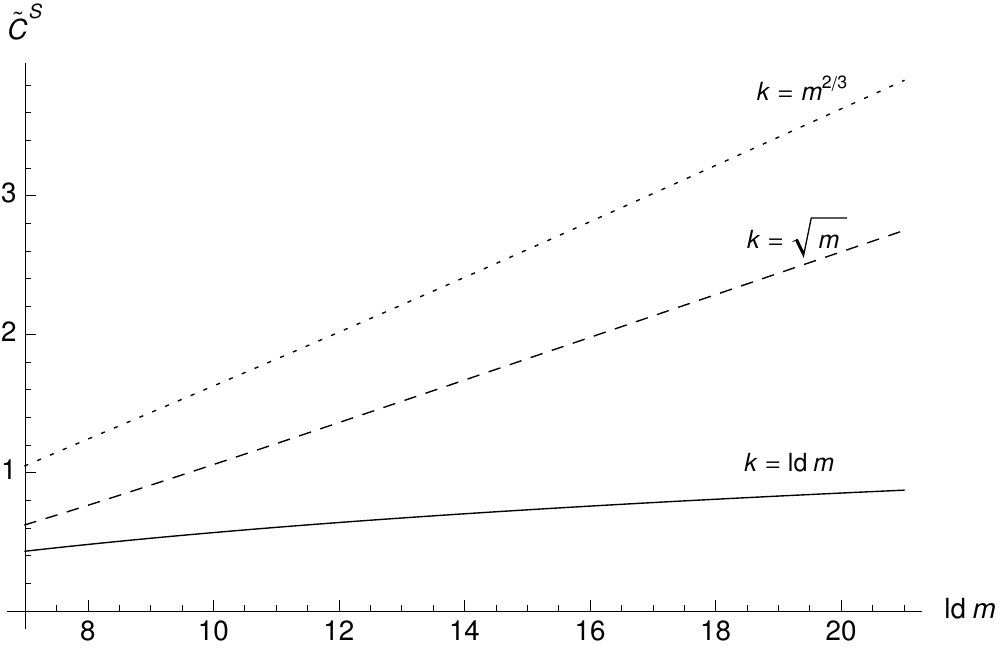}
\caption{Synaptic capacity $\tilde{C}^S$ (in bits per synapse) for the inhibitory Willshaw rule in the same conditions of figure \ref{fig:C-k-m-large}, calculated through normalisation of $C$ (cf. equation \ref{eq:network-capacity-2}) by $\tilde{F} \equiv 1 - p_1$. In the moderately-sparse coding regime (supra-logarithmic $k(m)$), which would otherwise lead to quickly vanishing $C$ and $C^S$ in the excitatory Willshaw model, the inhibitory network is capable of storing more than one bit per functional synapse already at surprisingly small $m$. As discussed in the main text, the synaptic capacity increases with $m$, as long as $k$ is as well an increasing function of $m$.\label{fig:CS}}
\end{center}
\end{figure}

What is remarkable is that as $m\to\infty$, the synaptic capacity $\tilde{C}^S$ diverges for any $k$ that increases with $m$, assuming that the binomial approximative theory we employ remains valid. For finite networks and activity levels of order $m^p$ with $0<p<1$, $\tilde{C}^S$ already surpasses unity for small- and medium-sized systems (see figure \ref{fig:CS}). Even for `classical' sparseness where $k$ is of logarithmic size, the capacity increases with network size (recall that $C$ was always vanishing for any non-constant $k$) and is always well above zero.

To picture the difference in capacities, for a network of size $m=10^6$, an error rate of $p_{01\epsilon}=0.01$ and a logarithmic activity level $k = \ln m \approx 14$, we obtain the network capacity $C \approx 0.03$, while the synaptic capacity is $\tilde{C}^S \approx 0.70$. If the coding level rises to a more realistic setting such as $k = \sqrt m = 1000$, the difference becomes drastic, as we have $C\approx 2.4\times 10^{-5}$ and $\tilde{C}^S \approx 2.6$.

There is a major qualitative change when the excitatory rule is replaced by the inhibitory one. Since $\tilde{F} \to 0$ as $k \to \infty$, in the limit of large networks the system is characterised by few synapses carrying a great amount of information. For moderate sparseness where $k$ is of the form $m^p$, $0<p<1$, and any setting of $p$, the synaptic capacity is (asymptotically)
\begin{equation}
\tilde{C}^S \approx 2p \left(- \ln p_{01\epsilon}\right)^{-1} \ln m,
\end{equation}
which grows with $m$ as fast as the corresponding asymptotic bound for the associative case \citep[see Table 1,][]{Knoblauch2010}, although here the high fidelity requirement enforced through the constant $p_{01\epsilon} > 0$ affects more strongly the obtained capacity. Note that the maximal pattern load is still large; substituting $k$ for $m^p$ in equation \ref{eq:Meps} we find
\begin{align}
\tilde{M}_\epsilon \approx -m^{2-2p} \ln \left(1-{p_{01\epsilon}}^{2m^{-2p}}\right),
\end{align}
which becomes, in the limit of large networks $m\to\infty$,
\begin{equation}
\label{eq:Mapprox}
\tilde{M}_\epsilon \approx 2p \cdot m^{2-2p} \cdot \ln m.
\end{equation}
When $k$ is of order $\sqrt{m}$, asymptotically we obtain the pattern capacity $\tilde{M}_\epsilon = m \ln m$, which is still supralinear in $m$, while the number of required functional synapses $\tilde{F}$ tends to zero.

In summary, considering that only functional synapses are relevant for the capacity measure, the Willshaw-type inhibitory learning rule leads to efficient familiarity discrimination in the limit of synaptic precision (two-state synapses). Interestingly, as in the pattern association task \citep{Knoblauch2010}, the network achieves high storage capacities for coding rates of the form $f=k/m=m^{p-1}=m^{-\alpha}$, $0<\alpha<1$, which for most cortical regions are (arguably) more realistic than the logarithmic levels required by the excitatory rule. If one accepts the logarithmic coding requirement, then the inhibitory model offers a pattern load that grows as $2 m^2 \mathop{\ln \ln} m \left(\ln m\right)^{-2}$ (see equation \ref{eq:Meps-limit}), still achieving capacities around one bit per synapse while maintaining high fidelity in the discriminator output and low anatomical connectivity.

\subsection{Corrections for binomially-distributed activity levels}
\label{sec:binomial}
To reach the former results we have assumed that the activity level per pattern was fixed at exactly $k$ firing neurons, at any given time, i.e., $\left|\tilde{\mathbf{x}}\right| = \left|\mathbf{x}^\mu\right|=k$ was kept constant across all $\mu$. Thus, all patterns were permutations of each other chosen from the $\binom{m}{k}$ possible configurations as in the analysis of \citet{Palm1980}. However, from the biological modelling perspective it might be more reasonable to take the assembly size as a random variable. In this section we let $\left|\mathbf{x}^\mu\right|$ and $\left|\tilde{\mathbf{x}}\right|$ assume a binomial distribution with characteristic probability $f \equiv k/m$, so that the mean activity level is still $k/m$, but the activity levels are allowed to vary.

In this case, the treatment is harder since we have to replace the constant parameter $k$ in the capacity analyses by a random variable. We denote by a star superscript `$^*$' whenever appropriate to differentiate quantities where $\left|\tilde{\mathbf{x}}\right|$ and $\left|\mathbf{x}^\mu\right|$ are random variables.

First, since the patterns have varying activity levels, to recover the `no-omission-errors' property $p_{10}=0$, we adjust the discrimination threshold for the excitatory network accordingly on a cue-by-cue basis,
\begin{equation}
\Theta_W^*(\tilde{\mathbf{x}}) = \left|\tilde{\mathbf{x}}\right|^2 = Z^2,
\end{equation}
denoting the binomially-distributed pattern activity level by random variable $Z$. The variable threshold could be implemented, alternatively, introducing an external feedforward inhibition field in the energy read-out, corresponding to a translation in the energy function,
\begin{equation}
H^*(\tilde{\mathbf{x}})=H(\tilde{\mathbf{x}})-\Theta^*_W(\tilde{\mathbf{x}}),
\end{equation}
implying $H^*(\tilde{\mathbf{x}})=0$ for familiar $\tilde{\mathbf{x}} \in \mathcal{S}$, as in the inhibitory Willshaw network implementation.

When the weights are set according to the inhibitory rule \eqref{eq:inhibitory-rule}, there is no need for the explicit external field, as the energy reads immediately $H(\tilde{\mathbf{x}}) = H^*(\tilde{\mathbf{x}})$ and the threshold can be simply set fixed $\Theta^*_I=\Theta_I=0$ as before. For the excitatory network, however, the variable threshold control is fundamental to stabilise the energy, as can be seen for instance through inspection of the variances of non-translated vs. translated energies (not shown here).

In the following, $p_B(x; n, p)=\binom{n}{x} p^x (1-p)^{n-x}$ is the probability mass function of the binomial distribution. We first approximate the conditional error probability by
\begin{align}
p_{01}^* &= \mathrm{P}(D(\tilde{\mathbf{x}}) = 1 \mid \tilde{\mathbf{x}} \notin \mathcal{S})\\
&\approx \sum_{i=0}^{M-1} p_B(i;M-1,f) \times \nonumber\\
&\times \sum_{z=1}^m p_B(z;m,f)\left(1-\left(1-f\right)^i\right)^{(z^2-z)/2},
\label{eq:p01-star-full}
\end{align}
which is the expression found by \citet{Buckingham1992} for the associative task under the same statistical assumptions, now adjusted to the quadratic familiarity discriminator; the full analysis of the distribution is due to \citet{Knoblauch2008}. Notice that equation \ref{eq:p01-star-full} is just an approximation, as the analyses of the associative case assume independence among the columns of $\mathbf{W}$. To compute the exact conditional error probability of the quadratic discriminator, however, would require analysing a $k \times k$ sub-matrix of $\mathbf{W}$, which is a difficult combinatorial problem we do not solve.

Approximating the exponent and employing the binomial approximation, as in \eqref{eq:p01}, we obtain
\begin{align}
\label{eq:p01-star}
p_{01}^* \approx \sum_{z=1}^m p_B(z; m, k/m) {p_1}^{z^2/2} \ge {p_1}^{k^2/2},
\end{align}
$p_1$ being the expected matrix load as given by \eqref{eq:p1}. Notice that in general, as expected and as in the case of the covariance rule \citep{Bogacz2002,Greve2009}, the error probability is never smaller than when the activity level is kept constant.

\begin{figure}[ht!]
\begin{center}
\includegraphics[scale=1.2]{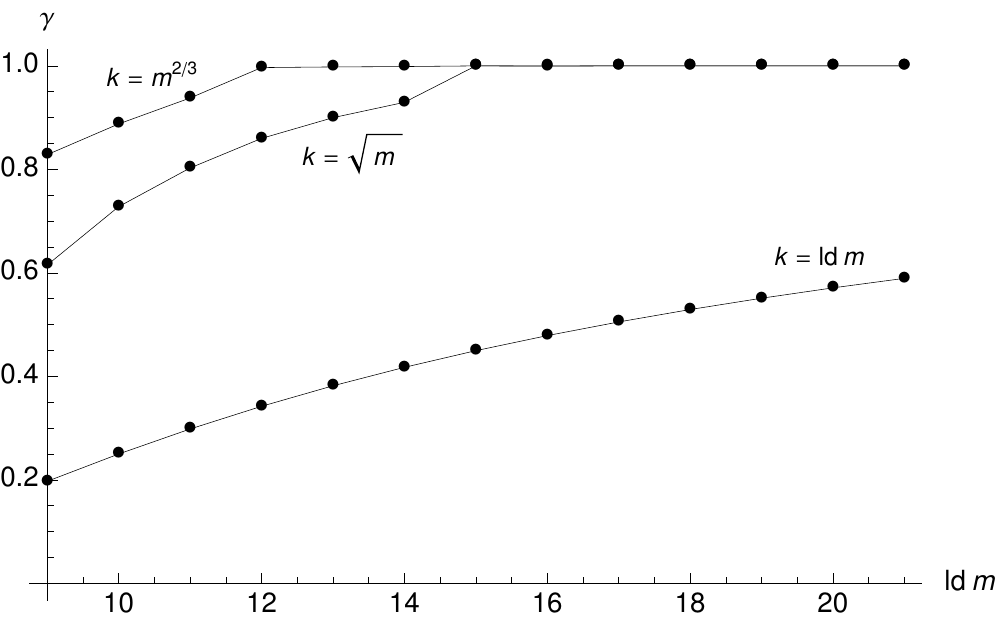}
\caption{The ratio $\gamma = \tilde{C}^{S*}/\tilde{C}^S$ between the obtained synaptic capacities (calculated through normalisation by $\tilde{F}$ of the network capacity of equation \ref{eq:network-capacity-2}) in the binomial- and fixed-activity pattern generation scenarios. Connecting (interpolating) lines are visual aids; solid markers represent the ratio of capacities computed for actual measured $M^*$ (binomially-distributed $Z$) vs. theoretical maximal $M_\epsilon$ (fixed $z=k$) as given by \eqref{eq:Meps}. The pattern load $M^*$ was found numerically by bisecting search over equation \ref{eq:p01-star} with the target $p_{01}^*$ set at $p_{01\epsilon}=0.01$. The relative difference between $C^{S*}$ and $C$ fades as $m$ grows and when the expected activity level order $k(m)$ increases.\label{fig:Cbin}}
\end{center}
\end{figure}

It is hard to obtain the pattern load $M^*$ as a function of $p_{01}^*$ without writing the summation in \eqref{eq:p01-star} in closed-form, which is difficult to accomplish due to the quadratic exponent. However, we can find numerically the $M^*$ such that the commission error probability $p_{01}^*$ is approximately equal to some arbitrary bound close to zero (say, $p_{01\epsilon}=0.01$), from which we compute the corresponding synaptic capacity $\tilde{C}^{S*}$. Then, to assess the impact of letting $k$ vary, we can see how the ratio $\gamma \equiv \tilde{C}^{S*}/\tilde{C}^S$ evolves as $m$ grows, for different mean activity levels.

As plotted in figure \ref{fig:Cbin}, $\gamma$ approaches unity as the network size parameter $m$ increases, and quickly so when the patterns are moderately sparse ($k = m^p$). For small, finite $m$ there is a rather large factor affecting $M^*$ that originates in the disorder introduced by the variability in the activity levels. This factor can be (approximately) as large as $1/5$ for $k$ of logarithmic size but attenuates as $m$ grows. Our numerical analysis strongly suggests then that the system remains qualitatively intact and the former conclusions drawn for fixed $k$ should hold, even for finite networks, although the discriminator is subject to a correcting factor which decreases the capacity of the model.

\section{Discussion}
If one restricts the model to operate with two-state synapses, a well-known and simple local update scheme can offer a surprising familiarity discrimination capacity, provided that the firing rates are kept low. We have analysed both the original Willshaw rule \citep{Willshaw1969} and a variation for inhibitory synapses recently proposed by \citet{Knoblauch2010}.

At high pattern loads, the traditional excitatory implementation imposes high connectivity and a heavy coding restriction; we have seen that for large enough networks the network capacity eventually approaches zero unless the activity levels are kept constant (independent of network size) and very low at all times. For neural populations of moderate size and low activity levels (e.g., of logarithmic order), one can obtain in the high-fidelity regime information and pattern capacities that are comparable to those found for the optimal linear rule. In this case, we find a rather low overall stored information content per synapse in comparison to the typical values achieved in the associative memory task, a fact that has already been discussed by \citet{Barrett2008,Greve2009}.

Taking into consideration that in the long-term the brain might prune silent synapses (that play a non-functional role and are mere spatial candidates for future potentiation) in stable memories and then place synapses in new locations as needed, \citet{Knoblauch2010} suggested the so-called synaptic capacity measure where only functional resources are taken into account. The critical observation we reach in our work is that the familiarity detection task parametrisation leads naturally to the dense potentiation regime, even for logarithmic sparse coding, which explains the large capacities achieved by the inhibitory Willshaw rule. In this case, we recover the increasing capacity function (with respect to network size) that is typical of the associative task.

Of course, another question altogether is to locate such structures in the actual central nervous system, and to ascertain if the less conservative inhibitory rule (where connections corresponding to previous coincidental activity are depressed and then pruned) is plausible and if it is actually observed in real synapses. It is worth noting that we have switched to an inhibitory circuit so that the energy `readout' mechanism \eqref{eq:discriminator} could remain intact, except for a change in the threshold. However, one could consider a sign-reversed connectivity matrix, i.e., an excitatory network implementation with exactly the same couplings as the inhibitory one. In this case, the less well-known inhibitory synaptic plasticity processes would be avoided, but the task would change, as a stronger excitatory signal would be elicited in the presence of novel patterns. Such a model could be appropriate to describe a novelty detection mechanism in regions where stronger excitatory activity is observed as a response to non-familiar stimuli. Our analysis should hold, as only the number (and not the type) of required functional synapses matters for the synaptic capacity measure we have considered.

Following the previous studies of familiarity detection, our analysis has focused on simple high-level modelling assumptions that could be refined if the biological implications require so. For instance, one could consider incorporating well-known features of more realistic or detailed models, such as stochastic synaptic transmission, arbitrary query noise, or spiking neurons.

\section*{Acknowledgements}
We would like to thank the two anonymous reviewers for detailed and thorough comments, as well as Francisco Burnay, \^{A}ngelo Cardoso, Francisco S. Melo and Diogo Rendeiro for carefully reading a preliminary version of the manuscript. This research was supported by the portuguese \textit{Funda\c{c}\~{a}o para a Ci\^{e}ncia e Tecnologia} (INESC-ID multiannual funding) through the PIDDAC Program funds and a doctoral grant awarded to the first author (contract SFRH/BD/66398/2009).

\appendix
\section{Derivation of the mutual information per pattern}
\label{apx:mutual-info}
For a given pattern transmission described by the true class (novel-familiar) of the pattern $X^\omega$ and the network output $Y^\omega$, we can define the mutual information $I(X^\omega; Y^\omega)$ in terms of the discriminator entropy $I(Y^\omega)$ and the conditional entropy $I(Y^\omega \mid X^\omega)$ of the discrimination outcome given the correct classification,
\begin{align}
\label{eq:single-mutual-info}
I(X^\omega; Y^\omega) &= I(Y^\omega) - I(Y^\omega \mid X^\omega).
\end{align}

Let us denote by $I(p) = -p \ld p - (1-p) \ld (1-p)$ the Shannon entropy in bits of a binary random variable $X$ with $\mathrm{P}(X=1)=p$ and $\mathrm{P}(X=0)=1-p$. Then, we can write the entropies in \eqref{eq:single-mutual-info} with respect to the prior probability $p \equiv P(X^\omega=1)$ and the error probabilities $p_{10}$ and $p_{01}$ \citep{Cover2006}, leading to
\begin{equation}
I(Y^\omega) = I(p(1-p_{10})+(1-p)p_{01}) = I(p+(1-p)p_{01}),
\end{equation}
and
\begin{equation}
I(Y^\omega \mid X^\omega)=pI(p_{10})+(1-p)I(p_{01})=(1-p)I(p_{01}),
\end{equation}
recalling that $p_{10}=0$ under the threshold setting \eqref{eq:ThetaW-familiar-energy}.

Inserting the expanded entropies into expression \ref{eq:single-mutual-info}, and substituting $p=1/2$ (the probability of a pattern being familiar), we obtain
\begin{align}
I(X^\omega; Y^\omega) &= I\left(\frac{1}{2}(1+p_{01})\right)-\frac{1}{2}I(p_{01})\\
&=1-\frac{1}{2}\Big((1+p_{01})\ld(1+p_{01})+(1-p_{01})\ld(1-p_{01})\Big)\nonumber\\
&-\frac{1}{2}\Big(-p_{01}\ld p_{01}-(1-p_{01})\ld(1-p_{01})\Big)\\
&=1-\frac{1}{2}\Big((1+p_{01})\ld(1+p_{01})-p_{01} \ld p_{01}\Big),
\end{align}
which is the expression presented in the main text (equation \ref{eq:network-capacity-2}).

\end{document}